# Friction force microscopy: a simple technique for identifying graphene on rough substrates and mapping the orientation of graphene grains on copper

**A J Marsden[1], M Phillips[2] and N R Wilson[1]**

[1] Department of Physics, University of Warwick, Coventry, CV4 7AL, UK

[2] Asylum Research UK Ltd., Commerce House, Telford Road, Bicester OX26 4LD, UK

Email: neil.wilson@warwick.ac.uk

**Abstract.** At a single atom thick, it is challenging to distinguish graphene from its substrate using conventional techniques. In this paper we show that friction force microscopy (FFM) is a simple and quick technique for identifying graphene on a range of samples, from growth substrates to rough insulators. We show that FFM is particularly effective for characterising graphene grown on copper where it can correlate the graphene growth to the three-dimensional surface topography and map the crystallographic orientation of the graphene nondestructively, reproducibly and at high resolution. We expect FFM to be similarly effective for studying graphene growth on other metal/locally crystalline substrates, including SiC, and for studying growth of other two-dimensional materials such as molybdenum disulphide and hexagonal boron nitride.

## 1. Introduction

Graphene is rapidly evolving from a material with fascinating fundamental properties to one with a wide range of applications. These applications involve integrating graphene onto, or into, other materials, such as dielectrics for electronic applications or polymers for functional composites. The range of applications is mirrored by the range of synthesis techniques. The original mechanical exfoliation has been supplemented by more scalable techniques such as sublimation of SiC, growth on transition metals, and liquid or chemical exfoliation [1]. In particular, chemical vapour deposition on copper foil [2] has been recently developed as an attractive approach for large area production of graphene for electronic and opto-electronic applications [1, 3].

This diversity of applications and synthesis routes creates a necessity for characterisation tools capable of identifying graphene on a range of substrates. Perhaps the simplest of such tools is optical microscopy; this has been used to great effect for identifying mono-layer graphene flakes on dielectrics such as silicon oxide. However, graphene is only visible on these substrates due to interference effects [4] placing strict requirements on the substrates and with the conventional limit of micron scale resolution of optical microscopy. Raman spectroscopy is often the method of choice for identifying graphene, particularly due to its ability to differentiate between monolayer, bilayer and few-layer graphene [5, 6]. However, mapping with Raman spectroscopy is extremely slow and the spatial resolution is again limited to around the micron level. Higher resolution analysis is possible with electron microscopy, for example scanning electron microscopy is routinely used to study graphene growth on copper with the observed contrast attributed to the low secondary electron emission of graphene[7]. SEM usually requires a conducting surface and image contrast is not readily interpreted to give a true representation of the three-dimensional surface topography. TEM has been extensively used for atomic-resolution imaging of graphene sheets [8], is capable of distinguishing between monolayer and bilayer graphene [8, 9], and has emerged as the most widely available technique for robustly determining the size and orientation of graphene grains in the continuous polycrystalline graphene sheets synthesized by CVD [10-12]. However, this type of TEM analysis requires the graphene to be removed from its substrate, precluding analysis of graphene in situ on substrates and complicating sample preparation.

In this work we show that friction force microscopy is a simple technique for identifying graphene even on rough insulating samples. It is particularly effective for analysing graphene grown on metal substrates, allowing the inter-relationship between surface topography and graphene growth to be observed. We demonstrate that FFM is capable of rapidly and reproducibly resolving the crystallographic orientation of graphene grown on copper, enabling graphene grain orientations to be mapped.

## 2. Experimental Methods

*2.1 Graphene Growth and Transfer*
Graphene was grown on copper foils via low pressure chemical vapour deposition (LP-CVD), as reported previously [13]. Copper foils (99.5% purity, 0.025 mm thick, Alfa Aesar product number 13382) were annealed for 20 minutes at 1000 $^{o}$C under vacuum ($4 \times 10^{-1}$ mbar) with hydrogen flowing. Methane was then introduced as the carbon feedstock for graphene growth, and the foils were cooled in a hydrogen atmosphere. Typically 20 standard cubic centimetres (sccm) of hydrogen with 5 sccm of methane for 30 seconds was used for a partial coverage of graphene, or 2 sccm hydrogen and 35 sccm methane for 10 minutes for a full coverage of graphene.

Graphene films were transferred onto silicon oxide using a polymer supported transfer process. A bilayer of methylmethacrylate (MMA) and poly methylmethacrylate (PMMA) was spin coated onto the foils, and the copper removed with copper etchant ($FeCl_3$, Alfa Aesar 44583). After etching, the graphene/MMA/PMMA was washed by repeated transfers to deionized water baths before it was scooped onto the silicon oxide substrate. Finally, the polymer supports were removed by soaking in acetone, followed by rinsing in isopropanol and blow-drying in nitrogen.

Graphene films were transferred on to PMMA beams using a similar polymer supported transfer. A single layer of PMMA was spin coated on the foil and then placed onto a PMMA beam, with a small drop of PMMA used to adhere the two surfaces. After curing in a vacuum furnace (1 mbar) at 60 $^{o}$C, the stack was floated copper side down in the copper etchant. Once the etch was complete, the PMMA beam was gently rinsed with DI water.

*2.2 AFM*
FFM images in Figures 1, 2 and 6, were taken on an Asylum Research MFP3D-SA AFM with Mikromasch NSC18 tips (nominal normal spring constant, resonance frequency and tip radius of 3.5 N/m, 75 kHz and 10 nm respectively). For quantitative information, both NSC18 and LS17 (nominal normal spring constant, resonance frequency and tip radius of 0.15 N/m, 12 kHz and 30 nm respectively) tips from Mikromasch were used. The tips were calibrated in the normal direction using the Sader method [14] and in the lateral direction using the wedge calibration method [15] on Mikromasch grating TGF11, which is an array of trapezoidal steps of known slope.

To resolve lattice information, i.e. Figures 3, 4 and 5, an Asylum Research Cypher was used with Mikromasch NSC18 tips. The image FFT, figure 4(b), was calculated using Gwyddion [16].

*2.3 SEM*
SEM images were taken on a Zeiss Supra 55-VP field emission gun SEM at an operating voltage of 10 kV using an in lens detector.

## 3. Results and discussion

*3.1 Distinguishing graphene from the copper surface*
Graphene grown on copper foil provides an interesting test for graphene characterisation tools. The graphene is not readily visible optically hence the conventional approach is to use SEM. This is quick but does not resolve the true three-dimensional structure of the surface, whilst recent reports have shown the importance of the surface topography for the growth and nucleation of graphene [13, 17].

Figure 1(a) shows a contact mode topography map of the copper foil; the surface is rough with the topography varying at the micrometre scale, dominated by parallel striations separated by around 10 μm that are formed by the cold-rolling process used to fabricate the foils [17]. With such large height variations it is not possible to discern the < 1 nm thick graphene sheets in the topography image. The deflection image, figure 1(b), shows roughness on the sub-micrometre scale and indicates the presence of islands which could be graphene, however, from this image alone it is not possible to say whether graphene is present or to differentiate areas of the copper surface covered in graphene from those which are bare. The contact mode image was taken with the scan angle perpendicular to the AFM cantilever, so that friction forces between the tip and surface result in lateral forces that twist the cantilever, measured as a lateral deflection. The lateral deflection image in the trace and retrace directions (i.e. as the tip is scanned to the right and to the left) are given in figures 1(c) and 1(d). Subtracting these two images gives a map that can be loosely interpreted as a local measure of the friction coefficient of the sample (see section 3.2 below), as shown in figure 1(e). This friction force microscopy (FFM) map clearly shows two different types of region on the sample: high friction regions (bright), and low friction regions (dark). This is emphasized by the line scans through the deflection images, marked as green and red lines in figures 1(a) and (c) and given in their respective colours in figure 1(f). As discussed below, the topography acts to increase or decrease the average lateral deflection, which explains the overall shape of the curves, whilst the difference between the two curves is a measure of the friction. The high friction and low friction regions are thus clearly visible in figure 1(f) and it is apparent that the magnitude of the friction is both consistent within each type of region and well separated between regions, allowing them to be clearly distinguished.

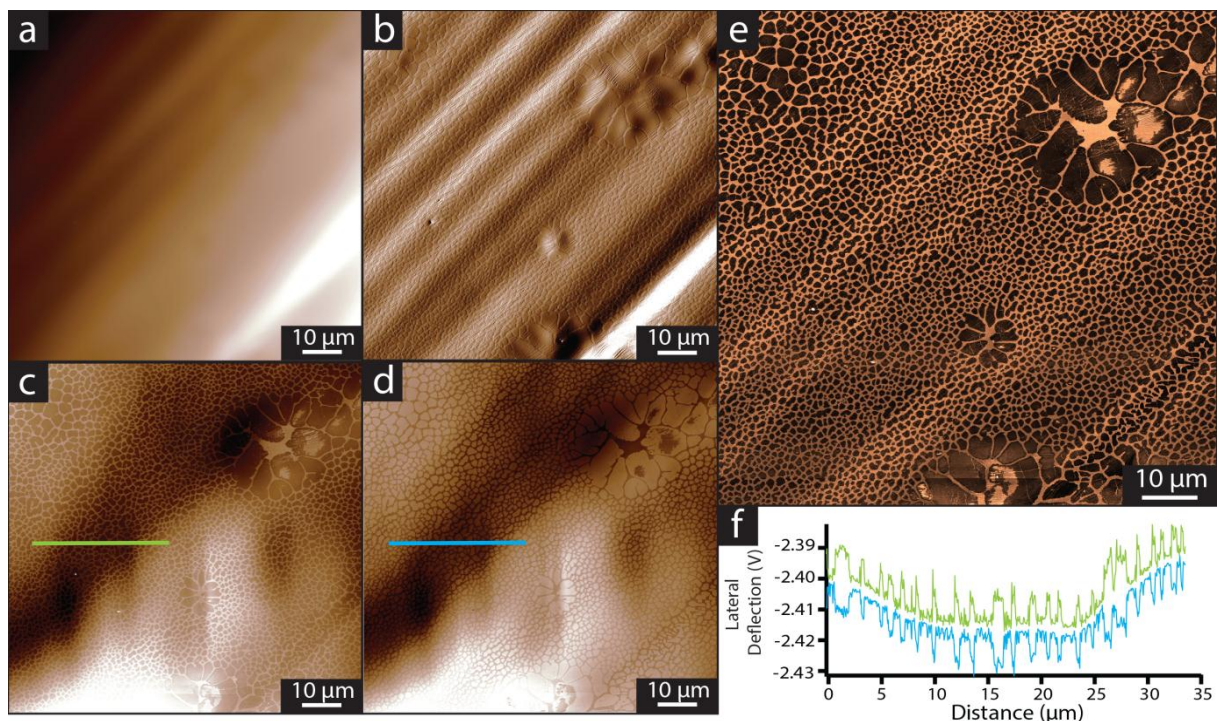

**Figure 1.** AFM of graphene on copper: (a) height image (full scale 6 µm), (b) deflection, (c) lateral deflection scanning left to right and (d) lateral deflection scanning right to left. A FFM map obtained from subtracting (d) from (c) is shown in (e). Line profiles from the green and red lines marked in (c) and (d) are shown in (f).

Figure 2 compares FFM with SEM of the same area of graphene grown on copper foil. The AFM topography image, Figure 2(a), shows an undulating surface with large height variations. The FFM image shows low friction (dark) and high friction (bright) regions as in Figure 1 (although note that the growth conditions were different for this sample). There is close correlation between the contrast in the FFM image with that of an SEM image of the same region, Figure 2(c). It is accepted that darker contrast on these samples in the SEM is due to graphene (which has a low secondary electron emission [7]), and the higher contrast due to copper. The low coefficient of friction (µ) of graphene has been well-documented, with previous reports showing that µ decreases from monolayer to bilayer to few layer graphene [18]; along with the direct comparison between FFM and SEM, this allows the low friction regions to be unambiguously associated with graphene, whilst the higher friction regions are bare copper. Darker regions are observed within the graphene islands in the SEM. Comparison with TEM has shown that these are due to additional graphene layers (i.e. bilayer or few layer graphene) [13]. There are some indications of different friction at these points, but not enough to definitively distinguish monolayer graphene from multilayer graphene. The FFM map has higher spatial resolution than the SEM image and more clearly distinguishes small contaminant particles from graphene and copper. It also allows direct comparison with the surface topography; for instance the surface undulation of the copper is barely visible, and certainly not quantifiable, in the SEM image.

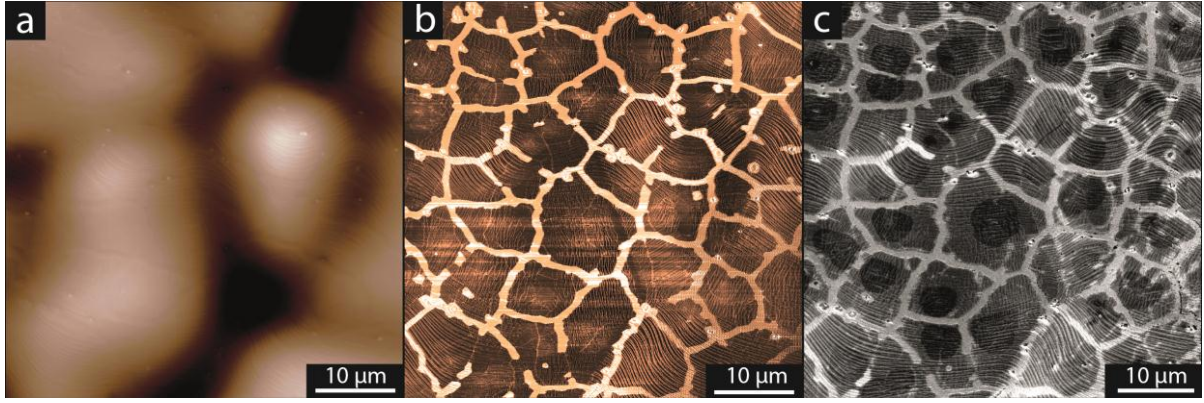

**Figure 2.** Comparison between AFM and SEM of graphene on copper: (a) height image (full scale 1.4 µm), (b) simultaneously acquired FFM map, (c) SEM image of the same area.

Clearly identifying both the surface topography and the graphene covered regions allows the two to be correlated; this enables the selective restructuring of the copper under the graphene to be resolved, as we reported recently [13]. Here we concentrate instead on the application of FFM to identify and map graphene.

*3.2 Quantitative analysis of the coefficient of friction.*
It is instructive to consider how the FFM signal is generated [15, 19]. The AFM tip is scanned with constant load ($L$), which results in a lateral friction force acting on the tip and hence a torsion ($T$) force acting on the cantilever. The torsion force applies a torsion moment to the cantilever, creating an angular deflection (or twist) of the cantilever. Most conventional AFMs, including the ones used here, use an optical lever method to detect the deflection of the cantilever, whereby a laser is reflected off the back of the cantilever onto a position sensitive photodetector (PSD) with four quadrants. The angular deflection, or twist, ($\phi$) of the cantilever results in a lateral movement of the laser spot on the PSD which is measured as a change in the lateral voltage output from the PSD ($\phi^0$). The torsion force ($T$) is linearly proportional to this lateral deflection, $\phi^0$ measured in volts,

$$T = \beta \phi^0$$
(1)

The constant of proportionality, $\beta$, depends on the optical lever sensitivity, the distance from tip to cantilever pivot axis, and the cantilever bending stiffness (or torsional spring constant) [15].

Assuming that Amonton's Law holds (i.e. that the lateral friction force is linearly proportional to the force normal to the surface), for a flat surface, the torsion force equals the friction force and is given by

$$T = \mu(L + A)$$
(2)

where µ is the coefficient of friction, $L$ the applied force and $A$ the adhesive force. However, on a sloped surface, the force normal to the surface (N) depends on the angle of the surface (θ) relative to the cantilever and the direction of motion. The torsion force then becomes

$$T_{\frac{u}{d}} = \frac{L \sin\theta \pm \mu(L\cos\theta + A)}{\cos\theta \mp \mu\sin\theta} = \pm\mu(L + A) + L\theta + \mu(L + A)\theta + O(\theta^2)$$

(3)

for a tip moving up a slope $T_u$, or down a slope $T_d$. It is clear from this equation that changes in the surface angle, $\theta$, (i.e. the surface topography) change the torsion force. The lateral signal in the AFM is coupled to the surface topography.

Subtracting the torsion force in either direction gives the width of the torsion loop (sometimes called the friction loop)

$$W = T_u - T_d = \frac{L\sin\theta + \mu(L\cos\theta + A)}{\cos\theta - \mu\sin\theta} - \frac{L\sin\theta - \mu(L\cos\theta + A)}{\cos\theta + \mu\sin\theta} = 2\mu(L+A) + O(\theta^2)$$

(4)

which has a much weaker dependence on the surface topography. This explains why for these rough surfaces the lateral voltage output, $\phi^0$, is correlated to the surface topography, whilst subtracting the data from the two scan directions gives a signal which reflects more accurately the local coefficient of friction. Experimentally it is the width of the lateral deflection loop (i.e. $W^0 = \phi_u^0 - \phi_d^0 = \beta W$) that is measured while the normal deflection (i.e. setpoint in contact mode) is varied, with calibration required to convert these signals to torsion loop, $W$, and load, $L$.

An example of this is shown in figure 3. FFM maps at two different loads are given in figures 3(a) and (c), note that the colour scale is the same in both images. Histograms of the images are given in figures 3(b) and (d) respectively. The histograms show two clearly resolved peaks, the lower one due to graphene and the higher due to copper. As the load increases the position of both peaks is shifted to higher friction values, but the friction force on the copper increases more than that on the graphene. Figure 3(e) plots the effect of load force (measured as the normal deflection) on the friction force (measured as $W^0$). The data are taken from a series of images such as those given in figures 3(a) and (c), with the points extracted by fitting the peaks in the histograms. Interestingly, figure 3(e) shows a linear relation between the load force and the friction force, i.e. confirming that a simple Amonton's law is obeyed. Previous work has suggested that for nanoscale contact areas, such as those realized in AFM, macroscopic laws of friction break down and deviations from a linear dependence of friction force on load can be observed [20]. However, we find no evidence for such behaviour here.

The coefficients of friction for graphene and copper can be extracted from the slopes of the lines in figure 2(e). This requires calibration of the cantilever normal spring constant and optical lever sensitivity and $\beta$ (equation 1). Here the Sader method was used to determine the normal spring constant [14] and the wedge calibration method [15] to calibrate $\beta$. Straight line fits to the data in Figure 2(e) gave µ = 0.7 ±0.2 for copper and µ = 0.18 ±0.05 for graphene. The uncertainties in the calibrations are large but repeated measurements using several different tips gave values in the range 0.12-0.18 for graphene. These values are roughly consistent with others in the literature: 0.03 for graphene on nickel [21], 0.22 for Cu-grown graphene transferred to $SiO_2$ [22].

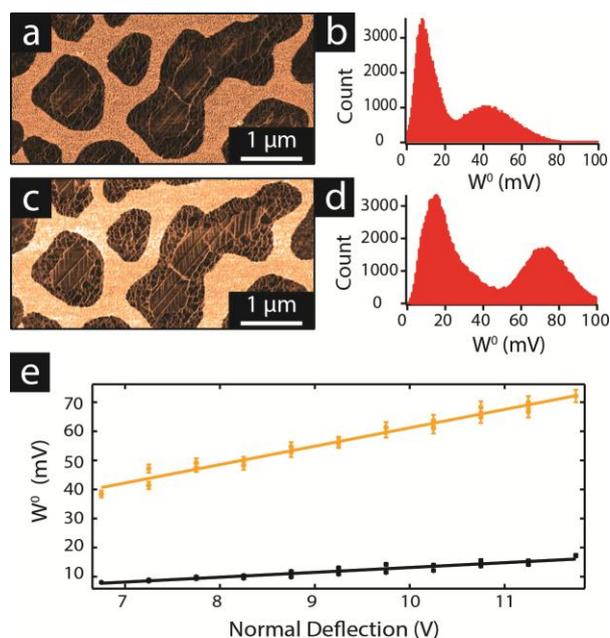

**Figure 3.** Effect of load on friction force: FFM maps, (a) and (c), and their corresponding histograms, (b) and (d) of graphene on copper taken with contact mode setpoints (i.e. normal deflection) of 6.8±0.1 V, (a), and 11.8±0.1 V, (c). (e) A graph showing the effect of varying load (through varying the normal deflection) on the friction force (measured as the width of the lateral deflection loop, $W^0$). The orange points are from copper and black from graphene, the lines are straight line fits to the data.

Importantly, it is not necessary to calibrate the FFM maps to identify graphene. Without calibration, FFM is a quick and simple technique to apply. Using cantilevers with comparatively long tips, such as the ones used here, increases the magnitude of the lateral deflection and hence increases the sensitivity. For rough surfaces where the topography adds a significant contribution to the lateral deflection signal, an AFM with closed loop scanning is beneficial to accurately subtract the trace and retrace lateral deflection signals.

*3.3 Resolving and mapping the orientation of the graphene lattice.*
Graphene grown on copper initially forms individual islands, as in figure 1, that, with further growth, merge to form a polycrystalline 'patchwork' quilt [10, 23]. The physical properties, e.g. mechanical strength and electrical conductivity, of the resultant graphene film are dependent on the grain boundaries between the graphene grains, the properties of which in turn depend on the relative orientations of the graphene grains [24, 25]. The standard technique so far for determining the sizes of the graphene grains and their relative orientations on the microscopic scale has been TEM [10-12], but this requires complex, time-consuming sample preparation that removes the graphene from the copper and hence loses information about the substrate topography. Alternative techniques are thus desirable. Yu et al. have shown that nematic liquid crystals (NLC) can orient themselves on graphite, enabling the orientation of the graphite to be mapped through analysing the director orientation of NLC deposited on graphene [26]. This offers the potential for large scale orientation mapping of graphene, but is comparatively low resolution (limited by the optical microscopy) and requires some sample preparation and expertise in NLC. For atomically flat surfaces, Lee et al. have shown that FFM can resolve the orientation of the graphene lattice [18] and Nemes-Incze et al. used AFM to deduce the crystallographic orientation of graphene grains transferred to atomically flat mica, noting that '*The ease with which atomic resolution AFM images can be acquired on graphene under ambient conditions is surprising*' [27]. Here we show that this can be extended to map the orientation of the graphene grains for CVD grown graphene in-situ on the copper growth substrate.

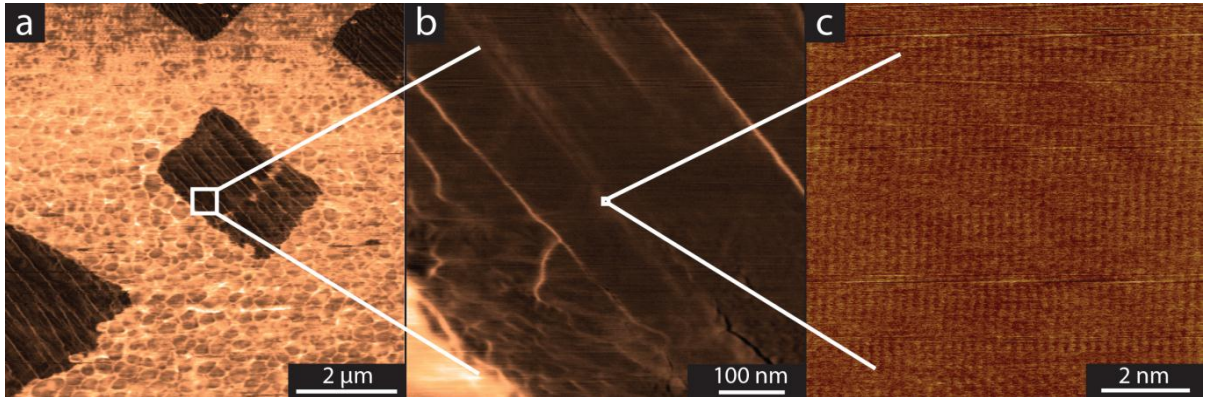

**Figure 4.** Lattice resolution: (a) and (b) are FFM images of graphene on copper, and (c) a lateral deflection map. The series from (a) to (c) were taken by zooming in on a graphene island with the area of the scan in (b) marked by the white box in (a), and (c) marked by the white box in (b).

Figure 4 shows a series of images of the same region at consecutively higher resolution[1]. As before, the graphene grains are readily distinguished from the copper surface. The copper surface is faceted under the graphene, so that on a given facet the surface is flat. When imaged at high resolution, figure 4(c), a periodic lattice is observed in the lateral deflection signal. This resolution is sufficient to determine the crystallographic orientation of the graphene grain. It should be noted that these high-resolution images were taken with a conventional commercial AFM tip under ambient conditions using a conventional commercial AFM and that the lattice resolution was immediately and readily apparent in the lateral deflection images but only occasionally in the topography or height images. The image is not showing atomic resolution; the contact area between the tip and sample is of order 100 nm$^2$, but the periodic stick-slip atomic scale forces are sufficient to generate a clear lattice resolution image.

The lattice orientation is more accurately measured on a larger image, through analysis of its fast Fourier transform (FFT), as shown in figure 5. In the lateral deflection image, figure 5(a), the graphene lattice can be resolved. An FFT of this image, figure 5(b), shows a clear hexagonal arrangement of peaks or spots (a diffractogram of the graphene lattice); the more intense peaks, marked by the dashed red hexagon in figure 5(b), correspond to the hk=10 type lattice vectors of graphene of length 1/0.213 nm$^{-1}$. The position of these spots enables the orientation of the graphene lattice to be readily measured[2]. Here the orientation is defined by the angle θ, measured arbitrarily relative to the 'x' (slow scan) direction of the image. This is the angle between $[10\bar{1}0]$ of graphene and the slow scan direction of the AFM tip.

---

[1] The high resolution image in (c) is shown as a lateral deflection map rather than a FFM map as small amounts of drift and hysteresis make it difficult to accurately subtract the lateral deflection in each direction at a given point
[2] Such FFT analysis is best performed on a relatively large scale image where the lattice corresponds to a few pixels.

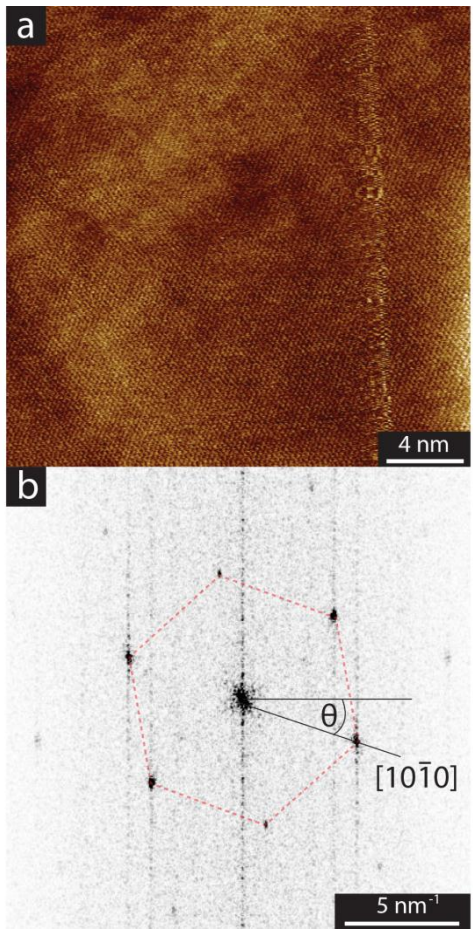

**Figure 5.** Determining the orientation of the graphene lattice: (a) lattice resolution lateral deflection image of graphene on copper. (b) FFT of (a), note that the contrast has been inverted for clarity such that the dark spots correspond to high intensity. Overlaid on (b) is a hexagon highlighting the spots due to the (*hk*) = (10) type graphene lattice vectors.

Lattice resolution images of graphene on copper are surprisingly easy to obtain reproducibly and at high scan rates (>20 Hz), making their acquisition comparatively rapid (< 30 seconds for a 512 line image). This makes it feasible to use this technique to map the orientations of graphene, as shown in figure 6.

Topography and FFM maps are shown in figures 6(a) and (b) respectively, with the isolated graphene grains again clearly resolved. A series of high resolution images were taken on a 16 by 16 grid spanning this area, their FFTs analysed and the angles of the graphene grains (as defined above) plotted as a colour scale image, figure 6(c). The empty (white) squares indicate images from which no orientation could be measured. In figure 6(d) the map of orientations is overlaid on the FFM map, with the positions of each image marked by white crosses.

There is a strong correlation between the points from which an orientation was not extracted (the white squares in figure 6(c)) and the copper regions. Of the 256 images, the FFM map shows that 74 are on the copper surface[3]. Of the 182 high resolution images on graphene, 175 gave a readily measurable orientation, a success rate of > 96%. The images were taken using an automated procedure, indicating that no special attention is required to achieve such a high success rate. This clearly demonstrates that lattice resolution FFM is a reproducible and reliable technique.

---

[3] This was also supported by the magnitude of the friction signal in the high resolution images.

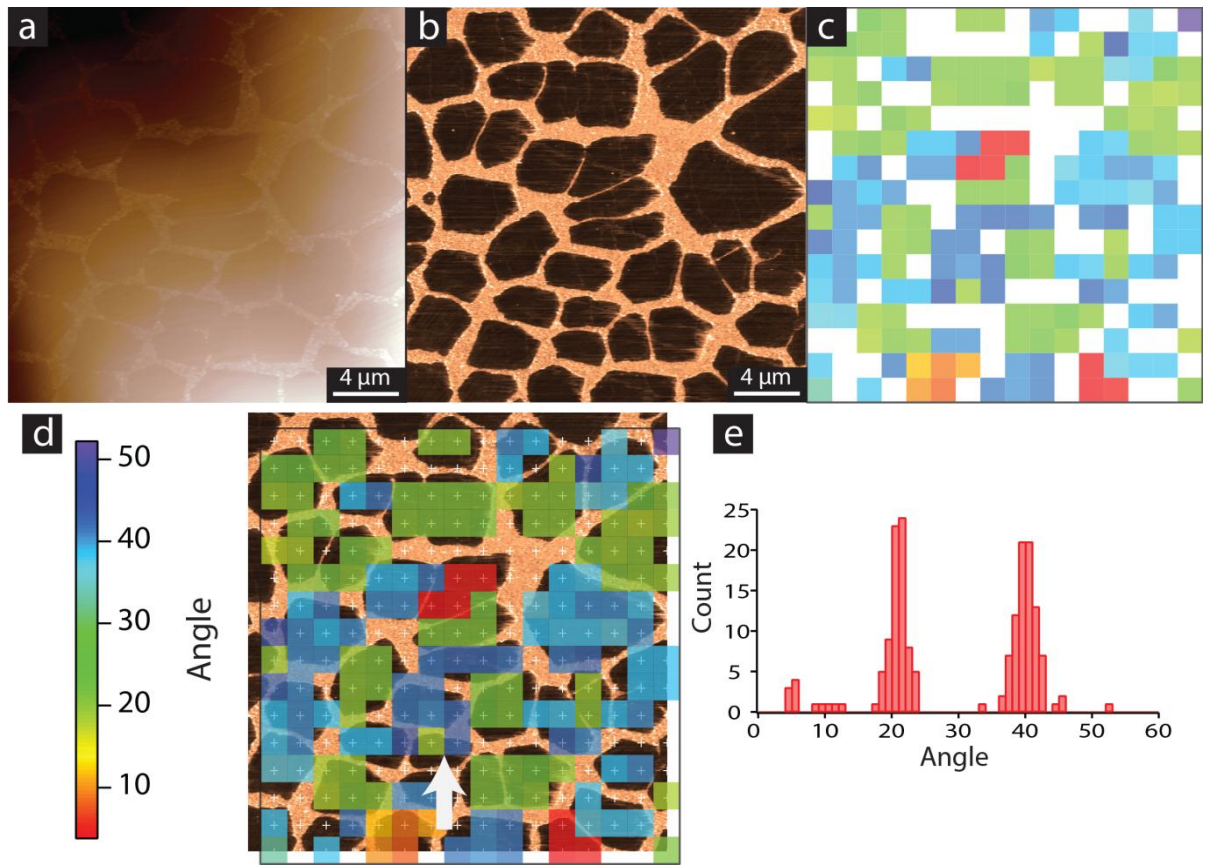

**Figure 6.** Graphene grain orientation mapping: (a) topography and (b) FFM of graphene grown on copper. (c) Orientation map of the corresponding area, taken from FFT analysis of a grid of images. The grid is marked by white crosses in (d) where the orientation map is overlaid on top of the FFM map; on the left is the color scale for the orientation map, the clear data points (white squares in (c)) denote images where no lattice was resolved. The white arrow indicates an island with a different orientation on either side. (e) Histogram of the orientation map shown in (c).

The data presented in figure 6 reveals important information about the graphene growth process. There are multiple measurements on most of the isolated graphene islands; in all but one of the grains the orientation is uniform across them, whilst in one (indicated by a white arrow) the orientation is different on either side. The individual islands have almost certainly nucleated and grown separately, the one island with two orientations suggests that coalescence can occur, even at relatively early stages, and that the graphene grains retain their orientation. Note that this information is not easy to obtain by TEM as that normally requires a continuous graphene film and loses all the information about the copper surface. By contrast, SEM can resolve the islands in situ on the copper, but not their orientations. Another important observation from figure 6(c) is that there are preferred orientations of the graphene grains. This is quantified in the histogram, figure 6(e), which shows two distinct peaks separated by 19±2 degrees. This is consistent with our previous electron diffraction studies [13] which revealed the importance of weak mismatch epitaxy in graphene growth. The orientation distribution found here by FFM mapping reinforces this observation and conclusively proves that it is due to isolated islands, separated by significant distances, nucleating and growing with the same orientation. This long range ordering must be due to the copper surface. The fact that graphene is nucleating with the same orientations points to the potential for another route to single crystal graphene on copper, beyond limiting nucleation, through the optimisation of the epitaxial nature of growth. FFM orientation mapping will be a crucial tool in such optimisation experiments.

*3.4 FFM identification of graphene on other substrates.*

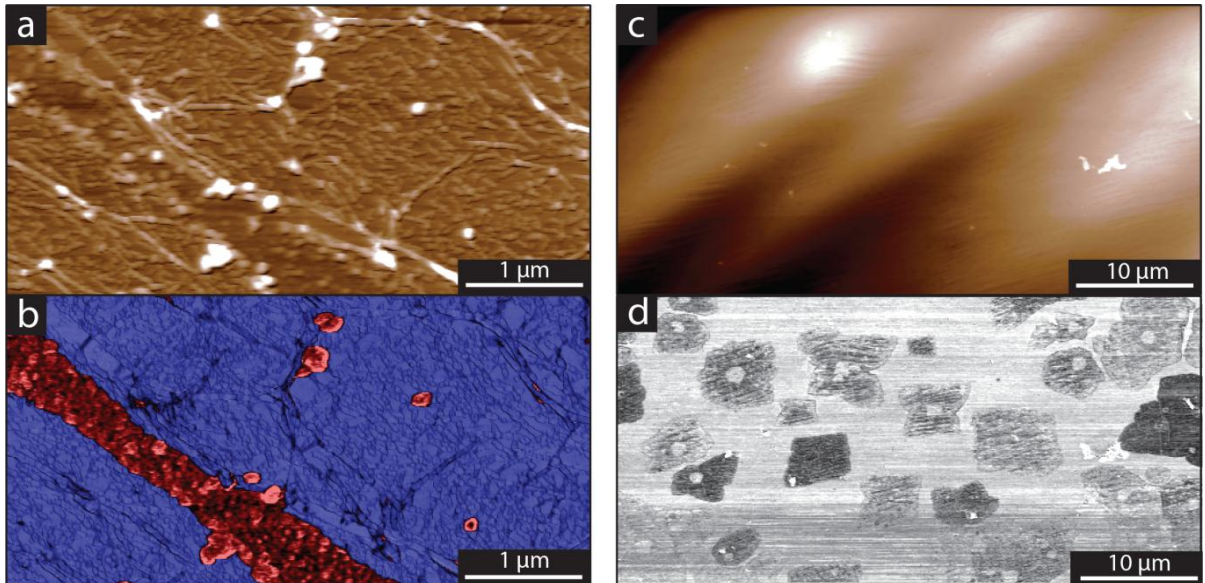

**Figure 7.** FFM of graphene on other substrates: (a) topography, 20 nm full height scale, and (b) FFM map of graphene transferred onto silicon oxide. In (b), blue corresponds to low friction and red high friction. (c) Topography, 800 nm full height scale, and (d) FFM map of graphene islands on PMMA.

FFM can also be used to identify graphene on other surfaces, including rough insulators. Figure 7(a) shows a topography map of graphene transferred to silicon oxide, as often used for testing the electrical properties of graphene. The transfer process uses a polymer support (as described in section 2.1) which is then dissolved. Remnants of the polymer layer remain, and the graphene can be creased or cracked by the transfer process. The simultaneously acquired FFM map, figure 7(b), clearly resolves the graphene regions (low friction, here the blue regions) from the polymer and silicon oxide (higher friction, here the red regions).

There is also much interest in combining graphene with polymers, for instance for mechanical reinforcement of composites. Figure 7(c) shows a topography map of graphene islands transferred on to a PMMA beam. The topography is fairly rough, making it impossible to identify the graphene islands in the topography image. They are, however, clearly resolved in the FFM image, figure 7(d), although currently we cannot explain the variations in contrast between islands or within the islands[4]. Note that the graphene on PMMA beam cannot be readily identified by optical microscopy, and as the sample is insulating it is not suited to SEM.

---

[4] Comparison with SEM of the graphene on copper before transfer suggests that the contrast within the islands is due to multi-layer regions, although that would suggest that here they have higher friction than the monolayer regions.

## 4. Conclusions

FFM is a simple and quick technique for identifying graphene on a range of samples, from growth substrates to rough insulators. It is particularly useful for studying graphene grown on copper where it can map the orientation of the graphene nondestructively, reproducibly and at high resolution. We expect FFM to be similarly effective for studying graphene growth on other metal/locally crystalline substrates. Previous work has shown that other two-dimensional crystals such as molybdenum disulphide and hexagonal boron nitride have low friction and can be readily imaged at lattice resolution by FFM [18], suggesting that this technique will be equally appropriate for studying such systems. The technique is standard on conventional commercial AFMs, is simple to apply, requires no sample preparation and offers complementary information to electron microscopy and Raman spectroscopy, as a result it has the potential to become a standard technique for graphene characterisation.

## Acknowledgements

We thank the EPSRC for support through a studentship for AJM. AJM and NRW acknowledge support from the Warwick Centre for Analytical Science (EP/F034210/1).